\documentclass[twocolumn,preprintnumbers,amsmath,amssymb]{revtex4}
\usepackage{amsmath,amsfonts,latexsym,amssymb,graphics,epsfig,subfigure,color,makeidx}
\usepackage{xcolor}
\usepackage{multirow}
\newcommand {\nn}    {\nonumber}

\begin{document} \large

\title{General  Covariant  Equations  for  Fields  of  Arbitrary Spin
\footnote{
         \large  Translated by E. J. Saletan\\}
\footnote{\large
         Commentary:
           In this paper, the Gel'fand-Iaglom field equation was generalized to the curved space based on the veilbein framework. The general covariant field equations for arbitrary spin were obtained. More importantly, in this work, Prof. Duan introduced the gauge covariant derivative with the local Lorentz transformation, which is the foundation for a gauge theory of gravitation. Therefore, this work is one of the earliest references to explore gauge theories of gravitation.
}
}

\author{Yi-Shi Duan}

\affiliation{\large{Joint Institute for Nuclear Research }}

\begin{abstract}\large
\begin{center}
          Submitted to JETP editor July 12, 1957\\
          J. Exptl.  Theoret. Phys.  (U.S.S.R.)  34, 632-636  (March, 1958)\\
          SOVIET  PHYSICS  JETP  VOLUME  34  (7),  NUMBER  3  SEPTEMBER,  1958 \\~\\
\end{center}
{The Gel'fand-Iaglom field equations are extended to the  general theory of relativity.}
\end{abstract}

\maketitle

To obtain a  generalized wave equation for a  field in general covariant form,  one  must define covariant differentiation of a  generalized wave function describing particles with arbitrary spin.  Gel'fand and Iaglom \cite{gelfand1}, Dirac \cite{dirac2},  and Fierz and Pauli \cite{fierz3}  have studied the generalized wave  equation in the  special theory of relativity.  In the present article, their theory is extended to the  general covariant form.

\section{ Semimetrics  and  semimetric  representation}

We  introduce the metric $g_{ik}$ in space-time with the  aid of the  asymmetric matrix $||\lambda_{i(\alpha)}||$ according to \cite{rumer4}
\begin{equation}\label{1}
g_{ik}=\lambda_{i(\alpha)}\lambda_{k(\alpha)},~||\lambda_{i(\alpha)}||^2=||g_{ik}||.
\end{equation}

Formally  $\lambda_{i(\alpha)}$  may be  thought of as half the metric  $g_{ik}$¡¤  We  shall call it the semimetric, and the representation shall be called the semimetric representation.  The metric  $g_{ik}$ remains invariant if the semimetric  $\lambda_{i(\alpha)}$ is subjected to the orthogonal transformation
\begin{equation}\label{2}
 \lambda'_{i(\alpha)}=L_{(\alpha\beta)}\lambda_{i(\beta)},
\end{equation}
where $||L_{(\alpha\beta)}||$ is an orthogonal matrix, which means that
\begin{equation}\label{3}
L_{(\alpha\beta)}L_{(\alpha\gamma)}=\delta_{(\beta\gamma)}.
\end{equation}
This is easily seen from  \eqref{2}  and \eqref{3},  according to which
\begin{equation}\label{4}
g'_{ik}=\lambda'_{i(\alpha)}\lambda'_{k(\alpha)}=\lambda_{i(\alpha)}\lambda_{k(\alpha)}=g_{ik}.
\end{equation}
Therefore all the equations  of the  general theory of relativity must remain invariant with respect to two transformation groups,  namely,  (a)  the group of general transformations of all coordinates of the form
\begin{equation}\label{5}
x'^{i}=f^{i}(x^1,x^2,x^3,x^4),
\end{equation}
and  (b)  the group of orthogonal transformations of the elements of the semimetric matrix
\begin{equation}\label{6}
\lambda'_{i(\alpha)}=L_{(\alpha,\beta)}\lambda_{i(\beta)}.
\end{equation}

According to Eq. \eqref{1}
\begin{equation}\label{7}
|\texttt{Det}~ \lambda_{i(\alpha)}|=+\sqrt{\texttt{Det}~g_{ik}}\neq0.
\end{equation}
Denoting  the  elements  of  the  inverse  matrix $||\lambda_{i(\alpha)}||^{-1}$  by  $\lambda^{i}_{(\alpha)}$,  we  have
\begin{equation}\label{8}
\lambda^{i}_{(\alpha)}\lambda_{i(\beta)}=\delta_{(\alpha\beta)};~~
\lambda^{i}_{(\alpha)}\lambda_{k(\alpha)}=\delta_{k}^{i}.
\end{equation}
We define
\begin{equation}\label{9}
dx_{(\alpha)}=\lambda_{i(\alpha)}dx^{i},dx^{i}=\lambda^{i}_{(\alpha)}dx_{(\alpha)}.
\end{equation}
We  shall call these new coordinates  $x_{(\alpha)}$  the semimetric coordinates.  From \eqref{8}  and  \eqref{1} it follows that
\begin{equation}\label{10}
ds^2=g_{ik}dx^{i}dx^{k}=dx_{(\alpha)}dx_{(\alpha)}.
\end{equation}
This means that in semimetric space the  element of length is given by a  normal quadratic form and is invariant under the linear transformation
\begin{equation}\label{11}
x'_{(\alpha)}=L_{(\alpha\beta)}x_{(\beta)}.
\end{equation}
From \eqref{9} we obtain
\begin{equation}\label{12}
\lambda_{i(\alpha)}=\partial x_{(\alpha)}/\partial x^{i},~~
\lambda^{i}_{(\alpha)}=\partial x^{i}/\partial x_{(\alpha)}.
\end{equation}
It follows from \eqref{11} that
\begin{equation}
\frac{\partial \varphi}{\partial x_{(\alpha)}}=
\frac{\partial \varphi}{\partial x^i}\frac{\partial x^i}{\partial x_{(\alpha)}}=
\lambda^i_{(\alpha)}\frac{\partial \varphi}{\partial x^i},\nn
\end{equation}
so that
\begin{equation}\label{13}
\lambda^i_{(\alpha)}\partial/\partial x^i\rightarrow\partial/\partial x_{(\alpha)}.
\end{equation}

\section{Covariant  derivative  of  a  generalized  field  function }
Let us introduce an  $n-$dimensional  matrix vector in semimetric space, whose  components  $L_{(\alpha)}$ form a  set of $n$  Hermitian matrices satisfying the condition
\begin{equation}\label{14}
[L_{(\alpha)},~I_{(\beta\gamma)}]_{-}=\delta_{(\alpha\beta)}L_{(\gamma)}
-\delta_{(\alpha\gamma)}L_{(\beta)},
\end{equation}
where $I_{(\alpha\beta)}$  is an infinitesimal operator of a representation of the  group of linear transformations of Eq.  \eqref{11}.  These infinitesimal operators satisfy the commutation rules
\begin{eqnarray}
[I_{(\alpha\beta)},~I_{(\gamma\delta)}]_{-}
 &=& \delta_{(\alpha\gamma)}I_{(\beta\delta)}
+\delta_{(\beta\delta)}I_{(\alpha\gamma)} \nonumber \\
&-& \delta_{(\alpha\delta)}I_{(\beta\gamma)}
-\delta_{(\beta\gamma)}I_{(\alpha\delta)}.\label{15}
\end{eqnarray}
We  shall denote the contravariant and covariant components of this  matrix vector in Riemannian
space by
\begin{equation}\label{16}
L^{i}=\lambda^{i}_{(\alpha)}L_{(\alpha)};~~L_{i}=\lambda_{i(\alpha)}L_{(\alpha)}.
\end{equation}
Writing
\begin{equation}
I_{ij}=\lambda_{i(\beta)}\lambda_{j(\gamma)}I_{(\beta\gamma)},\nn
\end{equation}
we can readily obtain from  \eqref{14}, \eqref{15},  and \eqref{1}  the relations
\begin{eqnarray}
 ~[L_{i}, I_{jk}]_{-} &=&  g_{ij}L_{k}-g_{ik}L_{j};\label{17} \\
 ~[I_{ij},I_{kl}]_{-}  &=&  g_{ik}I_{jl}+g_{jl}I_{ik}-g_{il}I_{jk}-g_{jk}I_{il}.\label{18}
\end{eqnarray}

Two complex generalized field functions $\psi$ and $\bar{\psi}$ are called adjoint functions  if the $n$  Hermitian forms
\begin{equation}
\bar{\psi}L_{(\alpha)}\psi\nn
\end{equation}
make up  a  vector in semimetric space.  Under transformations of group (a)
\begin{equation}\label{19}
\bar{\psi}'L_{(\alpha)}\psi'=\bar{\psi}L_{(\alpha)}\psi,
\end{equation}
the  components of these vectors remain invariant. Under the transformations of group  (b),  we have
\begin{equation}\label{20}
\bar{\psi}'L_{(\alpha)}\psi'=L_{(\alpha\beta)}(\psi L_{(\beta)}\psi).
\end{equation}
The  generalized wave functions transform among themselves according to
\begin{equation}\label{21}
\psi'=S\psi,~~\bar{\psi'}=\psi S^{-1},
\end{equation}
where  $S$ is  a  matrix which varies from point to point and  is related to  $||L_{(\alpha\beta)}||$  by
\begin{equation}\label{22}
SL_{(\alpha)}S^{-1}=L_{(\alpha\beta)}L_{(\beta)},
\end{equation}
which follows  from  \eqref{20}.  In order to derive the covariant differentiation formula for  a  generalized field function,  we  must define the  concept of parallel displacement.  If two points  $x$ and  $x+dx$  are separated by an infinitesimal distance,  the wave functions  at these points  are related by the  infinitesimal linear transformations
\begin{eqnarray}\label{23}
\psi(x+dx)&=&[I+\Lambda_{i}dx^{i}]\psi(x),\nn \\
\bar{\psi}(x+dx)&=&\bar{\psi}(x)[I-\Lambda_{i}dx^{i}],
\end{eqnarray}
where $\Lambda$  is  a  certain matrix.

If \eqref{23}  is to define parallel displacement, the vector  $\bar{\psi}L_{(\alpha)}\psi$,  which is constructed of  $\bar{\psi}$ and $\psi$ , must undergo a  parallel displacement, which means that
\begin{equation}\label{24}
A_{\alpha}(x+dx)=\{\delta_{(\alpha\beta)}+\eta_{\sigma(\alpha\beta)}dx^{\sigma}\}A_{(\beta)}(x).
\end{equation}
From \eqref{23} and  \eqref{24}  it follows that
\begin{eqnarray}\label{25}
&\bar{\psi}(x)\big[I-\Lambda_{i}dx^{i}\big]L_{(\alpha)}\big[I+\Lambda_{i}dx^{i}\big]\psi(x) \nonumber \\
&=\bar{\psi}L_{(\beta)}\psi \big[\delta_{(\alpha\beta)}+\eta_{i(\alpha\beta)}dx^{i}\big],
\end{eqnarray}
where the  $\eta_{i(\alpha\beta)}$  are the  components  of the affine  connection, which have been given by Rumer \cite{rumer4}.

According to Eq.  \eqref{25},  the $\Lambda_{i}$  are given by
\begin{equation}\label{26}
L_{(\alpha)}\Lambda_{i}-\Lambda_{i}L_{(\alpha)}=\eta_{i(\alpha\beta)}L_{(\beta)}.
\end{equation}
Multiplying \eqref{26}  by $\lambda^{i}_{(\delta)}$ we obtain
\begin{equation}\label{27}
L_{(\alpha)}\Lambda_{(\delta)}-\Lambda_{(\delta)}L_{(\alpha)}=\eta_{(\delta\alpha\beta)}L_{(\beta)},
\end{equation}
where $\Lambda_{(\delta)}=\lambda^{i}_{(\delta)}\Lambda_{i}$, and the
$\eta_{(\alpha\beta\gamma)}=\lambda^{i}_{(\alpha)}\eta_{i(\beta\gamma)}$  are the Ricci curvature coefficients \cite{eisenhart5}.

The general solution of Eq.  \eqref{27}  will be
\begin{equation}\label{28}
\Lambda_{i}=\frac{1}{2}\eta_{i(\alpha\beta)}I_{(\alpha\beta)}+if_{i}I,
\end{equation}
where the  $f_{i}$ are arbitrary functions.  This  is easily seen by making use of \eqref{14}.

Writing $f_{(\alpha)}=\lambda^{i}_{(\alpha)}f_{i}$,  we obtain
\begin{equation}\label{29}
\Lambda_{(\delta)}=\frac{1}{2}\eta_{(\delta\alpha\beta)}I_{(\alpha\beta)}+if_{(\delta)}I.
\end{equation}
Thus the covariant derivative of a  generalized field function will be
\begin{eqnarray}
\begin{array}{c}
  \nabla_{i}\psi=\frac{\partial\psi}{\partial x^{i}}-\Lambda_{i}\psi, \\~\\
  \nabla_{i}\bar{\psi}=\frac{\partial\bar{\psi}}{\partial x^{i}}+\bar{\psi}\Lambda_{i}
\end{array}
 \label{30}
\end{eqnarray}
and in semimetric space
\begin{eqnarray}
\begin{array}{c}
  \nabla_{(\alpha)}\psi=\frac{\partial\psi}{\partial x_{(\alpha)}} -\Lambda_{(\alpha)}\psi, \\~\\
  \nabla_{(\alpha)}\bar{\psi}=\frac{\partial\bar{\psi}}{\partial x_{(\alpha)}} +\bar{\psi}\Lambda_{(\alpha)}.
\end{array}
 \label{31}
\end{eqnarray}

\section{ Extension  of  the  gel'fand-iaglom field  equations  to  the  general theory  of  relativity  }
The  covariant field equations for arbitrary spin were obtained in the  special theory of relativity by Gel'fand and Iaglom \cite{gelfand1}, and are of the form
\begin{equation}\label{32}
L^{k}\frac{\partial\psi}{\partial x^{k}}+m\psi=0,
\end{equation}
where $\psi$  is a  generalized field function describing particles with arbitrary spin, and the  $L^{k}$  are matrices which determine the  linear transformation properties of the  $\psi$  function.

In order that Eqs. \eqref{32}  become covariant with respect to all physically possible transformations, the ordinary derivatives $\partial\psi/\partial x^{k}$  which appear in them must be replaced by covariant derivatives $\nabla_{k}\psi$.  Then the  general covariant field equations will be
\begin{equation}\label{33}
L^{k}\nabla_{k}\psi+m\psi=0,
\end{equation}
Inserting \eqref{30}  into \eqref{33}  and using \eqref{28},  we  obtain
\begin{equation}\label{34}
L^{k}\frac{\partial\psi}{\partial x^{k}} +m\psi-\frac{1}{2}L^{k}\eta_{k(\beta\gamma)}I_{(\beta\gamma)}\psi=0,
\end{equation}
where the  $L^{k}$ are matrix functions  satisfying Eqs. \eqref{17}.  From \eqref{13}, \eqref{16},  and \eqref{34} one  can obtain the general covariant field equations in semimetric space, namely
\begin{equation}\label{35}
L_{(\alpha)}\frac{ \partial\psi}{\partial  x_{(\alpha)}}+m\psi-\frac{1}{2}\eta_{(\alpha\beta\gamma)}L_{(\alpha)}I_{(\beta\gamma)}\psi=0,
\end{equation}
where the  $L_{(\alpha)}$  satisfy relations \eqref{14}.  If  $L_{(\alpha)}=\gamma_{(\alpha)}$, where the $\gamma_{(\alpha)}$ are Dirac matrices, then
\begin{equation}\label{36}
I_{(\alpha\beta)}=\frac{1}{2}[\gamma_{(\alpha)}\gamma_{(\beta)}-\gamma_{(\beta)}\gamma_{(\alpha)}],
\end{equation}
and \eqref{34}  become the general covariant Dirac equation
\begin{equation}\label{37}
\gamma_{(\alpha)}\frac{\partial\psi}{\partial x_{(\alpha)}} + m\psi-\frac{1}{4}\gamma_{(\alpha)}\gamma_{(\beta)}
\gamma_{(\gamma)}\eta_{(\alpha\beta\gamma)}\psi=0,
\end{equation}
a  special case of \eqref{34}  which has been previously obtained by Fock and Ivanenko \cite{fock6}.  This shows that Eqs. \eqref{34} and \eqref{35}  are of greater generality.  The general covariant Lagrangian is in this case
\begin{eqnarray}
L&=&\frac{1}{2}\left\{\bar{\psi}L^{k}\Big(\frac{\partial\psi}{\partial x^{k}}\!-\!\Lambda_{k}\psi\Big) \right. \nonumber \\
 && \left.~~~~ -\Big(\frac{\partial\bar{\psi}}{\partial x^{k}}\!+\!\Lambda_{k}\bar{\psi}\Big)L^{k}\psi\!+\!2m\bar{\psi}\psi\right \}.\label{38}
\end{eqnarray}
If \eqref{38}  is substituted into Euler's equation,  one easily obtains Eq. \eqref{34}.  Further,  it is easily shown that the  symmetric energy-momentum tensor and the current vector in general covariant form can be written,  respectively,
\begin{eqnarray}
\label{3940}
T_{ik}&=&\frac{1}{2}\Big[\bar{\psi}L_{k}\nabla_{i}\psi+\bar{\psi}L_{i}\nabla_{k}\psi\nn\\
&&~~-\nabla_{i}\bar{\psi}L_{k}\psi-\nabla_{k}\bar{\psi}L_{i}\psi\Big], \\
j^{k} &=& ie\bar{\psi}L^{k}\psi.
\end{eqnarray}

It should be noted that the field equations \eqref{35}, as opposed to those of the special theory of relativity, contain the additional operator terms
\begin{equation}
\delta\hat{m}=\frac{1}{2}\eta_{(\alpha\beta\gamma)} L_{(\alpha)}I_{(\beta\gamma)},\nn
\end{equation}
acting on the  spinor or tensor fields.  These operators may therefore be treated as mass operators entering the theory in a  natural way.  They are not introduced artificially or without any reasonable basis,  as is done in many works on ordinary quantum field theory.  One may hope that these operators will make  it possible to eliminate
the difficulties  associated with divergences in field theory.

In conclusion, I  express my gratitude to Professor Iu. M. Shirokov for discussing the results of the present work.


\begin{thebibliography}{100}\large
\bibitem{gelfand1}
M. Gel'fand  and  A. D. Iaglom,  J.  Exptl. Theoret. Phys.  (U.S.S.R.)  \textbf{18},  703  (1948).

 \bibitem{dirac2}
P. A. M. Dirac,  Proc.  Roy.  Soc.  (London) \textbf{A155},  117  (1939)  [sic! Probably ... \textbf{A155},  447
(1936) ].

 \bibitem{fierz3}
 M.  Fierz and W.  Pauli,  Proc.  Roy.  Soc. (London)  \textbf{A173},  211 (1939).

 \bibitem{rumer4}
Iu. Rumer,  J. Exptl. Theoret. Phys. (U.S.S.R.)  \textbf{2},  271  (1953).

  \bibitem{eisenhart5}
 L. Eisenhart,  Riemannian Geometry (Russ. Transl.),  1948.

  \bibitem{fock6}
V. A. Fock and D. D. Ivanenko,  Z. Physik  \textbf{30}, 678  (1929).
\end{thebibliography}
\end{document}